\documentclass[twoside,10pt,a4paper]{article}
\pdfoutput=1
\usepackage{graphicx}
\usepackage{xcolor}
\usepackage{amsfonts}
\usepackage{amsmath}
\usepackage{multirow}
\usepackage{array}
\usepackage[latin2]{inputenc}
\usepackage[OT4]{fontenc}
\usepackage[hang,raggedright]{subfigure}
\usepackage[matrix,frame,arrow]{xy}

\newcommand{\C}{\mathbb{C}}

\newcommand{\bra}[1]{\langle{#1}|}
\newcommand{\ket}[1]{|{#1}\rangle}

\renewcommand{\1}{\mbox{$\mathbb{I}$}}

\newcommand{\ketbra}[2]{\ket{#1}\bra{#2}}

\newcommand{\Tr}[1]{\mathrm{Tr}\left(#1\right)}

\newcommand{\qw}[1][-1]{\ar @{-} [0,#1]}

\newcommand{\gate}[1]{*{\xy *+<.6em>{#1};p\save+LU;+RU **\dir{-}\restore\save+RU;+RD **\dir{-}\restore\save+RD;+LD **\dir{-}\restore\POS+LD;+LU **\dir{-}\endxy} \qw}
\newcommand{\meter}{\gate{\xy *!<0em,1.1em>h\cir<1.1em>{ur_dr},!U-<0em,.4em>;p+<.5em,.9em> **h\dir{-} \POS <-.6em,.4em> *{},<.6em,-.4em> *{} \endxy}}

\newcommand{\multigate}[2]{*+<1em,.9em>{\hphantom{#2}} \qw \POS[0,0].[#1,0];p !C *{#2},p \save+LU;+RU **\dir{-}\restore\save+RU;+RD **\dir{-}\restore\save+RD;+LD **\dir{-}\restore\save+LD;+LU **\dir{-}\restore}
\newcommand{\ghost}[1]{*+<1em,.9em>{\hphantom{#1}} \qw}

\newcommand{\lstick}[1]{*!R!<.5em,0em>=<0em>{#1}}

\newcommand{\Qcircuit}[1][0em]{\xymatrix @*[o] @*=<#1>}

\begin{document}

\title{Noisy quantum Monty Hall game}

\author{Piotr Gawron\\
The Institute of Theoretical and Applied Informatics\\
of the Polish Academy of Sciences,\\
Ba\l{}tycka 5, 44-100 Gliwice, Poland\\
gawron@iitis.pl
}
\date{November 2, 2009}

\maketitle


\pagestyle{myheadings}


\begin{abstract}
The influence of spontaneous emission channel and generalized Pauli channel on
quantum Monty Hall Game is analysed. The scheme of Flittney and Abbott is
reformulated using the formalism of density matrices. Optimal classical
strategies for given quantum strategies are found. The whole presented scheme
illustrates how quantum noise may change the odds of a~zero-sum game.
\end{abstract}

\section{Introduction}
Game theory studies decision making for a~given set of rules, in order to select
a strategy to maximize one's pay-off. This theory is widely used in economics,
biology, sociology and sometimes in politics. Quantum game theory, the subclass
of game theory that involves quantum phenomena, lies at the crossroads of
physics, quantum information processing, computer and natural sciences
\cite{EWL1999,PS2003}.

The Monty Hall game has its roots in the popular TV show \emph{Let's Make a
Deal} and it was often the source of misunderstanding. There are several
quantisations of this game \cite{LZHG2001, FA2002, AGKKMW2002, ZCPP2006}. We
follow the scheme by Flitney and Abbott and reformulate it in the language of
the density matrices in order to study the influence of noise on the game
behaviour.

Problem of noise in quantum games was also studied in \cite{CAKKL2003,
Johnson2001, FH2007, Meyer2003, OSI2004, CKO2002, NT2006, GMS2008}.

This paper is organised as follows. In Section~\ref{sec:classical:MH} the
classical Monty Hall problem is introduced. In Section~\ref{sec:quantum:MH}
Flitney and Abott's scheme of quantisation of Monty Hall game is presented. In
Section~\ref{sec:noisemodel} the noise model which is applied to the game is
discussed. At last in Section~\ref{sec:results} the results and their discussion
is presented.

\section{Classical Monty Hall game}
\label{sec:classical:MH}
In the classical scheme Monty Hall game runs as follows.
There are two players: Alice and Bob. 
Bob's goal is to get the prize and Alice plays the role of banker.

There are three boxes of which only one contains the prize. 
The game consists of successive steps:
\begin{enumerate}
 \item Alice randomly chooses one box and hides the prize in it.
 \item Bob chooses one of the boxes according to his will.
 \item Alice opens one of the boxes which does not contain the prize.
 \item Bob now have an option to keep his choice or to switch and chose the 
 other closed box.
 \item Alice opens the box chosen by Bob.
\end{enumerate}
Bob wins if the prize is in the box he have chosen. Otherwise he looses.

The game is appealing and thought-provoking because Bob's optimal strategy
differs from intuition. To achieve higher probability of winning, Bob should
switch the box fourth step of the game. Explanation of this fact is very simple.
There are two possible cases in step two: Bob chooses the box with the prize
inside (with probability $\frac{1}{3}$) or Bob chooses the box without a~prize
inside (with probability $\frac{2}{3}$). Then in third step, when Alice opens
one of the boxes, Bob switches. Hence in the first case he will lose, but in the
second case he will always switch to the box containing the prize. Therefore
switching strategy yields to expected pay-off of $\frac{2}{3}$ and not switching
strategy only to $\frac{1}{3}$.

\section{Quantum Monty Hall game}
\label{sec:quantum:MH}
Flitney and Abbott presented in \cite{FA2002} following quantisation of this
game. Alice's and Bob's choices are represented by qutrits and the game starts
in some initial state which will be specified further. Players' strategies are
represented by operators acting on their respective qutrits. Third qutrit
represents the box opened by Alice.

The state of the system may be expressed by the normalised state vector
\begin{equation}
 \ket{\Psi}=\ket{o}\otimes\ket{b}\otimes\ket{a} \in \C^{3}\otimes\C^{3}\otimes\C^{3},
\end{equation}
where $a$ is Alice's choice, $b$ is Bob's choice and $o$ represents the box that
has been opened. 

The operator for opening of the box is defined as
\begin{equation} 
 {O}=\sum_{i,j,k,l=0}^2 |\epsilon_{i,j,k}|\ket{njk}\bra{ljk}+\sum_{jl=0}^2\ket{mjj}\bra{jll},
\end{equation}
and the operator for switching the box as
\begin{equation} 
 {S}=\sum_{i,j,k,l=0}^2 |\epsilon_{i,l,k}|\ket{ijk}\bra{ljk}+\sum_{ij=0}^2\ket{iij}\bra{iij},
\end{equation}
where $|\epsilon_{i,l,k}|=1 \Leftrightarrow i\neq l \wedge l\neq k \wedge i\neq k$, otherwise $|\epsilon_{i,l,k}|=0$,
$n=(i + l)\mod 3$ and $m=(j + l + 1)\mod 3$.

Bob's not switching operator ${N}$ is represented by the identity matrix acting
on the state of three qutrits.

Alice and Bob are restricted to unitary transformations on their qutrits. ${A},
{B} \in SU(3)$ are players' movements.

The unitary operator that implements this game is given by relation
\begin{equation} 
{G}_s={S}\cdot{O}\cdot({I}\otimes{B}\otimes{A}),
\end{equation}
if Bob chooses to switch or by relation
\begin{equation} 
{G}_n={N}\cdot{O}\cdot({I}\otimes{B}\otimes{A}),
\end{equation}
if Bob chooses not to switch the box.

The final state of the game is $\rho_{x}=G_x\rho_i G_x^\dagger$, where
$x\in\{s,n\}$ indicates Bob's strategy and $\rho_i$ is the initial state of the
game.

One may consider Bob's classic probabilistic strategies. Bob controls a~free
parameter $\gamma\in \left[0,\pi/2\right]$, which represents the mixing of
switching and not switching strategies. Pure switching strategy is obtained for
$\gamma=\pi/2$ and pure not switching strategy is obtained for $\gamma=0$.

Bob wins if he picks the correct box, hence expectation value of his pay-off is
given by the equation
\begin{equation} 
\langle\$_B \rangle=\sum_{ij=0}^2 \Tr{\ketbra{ijj}{ijj}\left((\cos{\gamma})^2\rho_{s}+(\sin{\gamma})^2\rho_{n}\right)}.
\end{equation}

Flitney and Abbott considered two initial states, one separable
\begin{equation}
\label{equ:psi1}
\ket{\Psi_{1}}=\ket{0}\otimes\frac{1}{\sqrt{3}}(\ket{0}+\ket{1}+\ket{2})\otimes\frac{1}{\sqrt{3}}(\ket{0}+\ket{1}+\ket{2})
\end{equation}
and one having qutrits of Alice and Bob entangled
\begin{equation}
\label{equ:psi2}
\ket{\Psi_{2}}=\ket{0}\otimes\frac{1}{\sqrt{3}}(\ket{00}+\ket{11}+\ket{22}).
\end{equation}

The goal of the this work is to analyse the influence of noise on the game
outcome. Therefore we assume, that the game is not played on the pure state but
on a~mixed state, that underwent non-unitary evolution trough a noisy channel.
 
The initial state of the game is given by
$\rho_i=\Phi(\ketbra{\psi_i}{\psi_i})$, where $\Phi(\cdot)$ denotes the quantum
noisy channel.

\section{Noise model}
\label{sec:noisemodel}
In this case, quantum Monty Hall game is implemented on qutrits, i.e. three
level quantum states. One can imagine that such a game could be implemented in
real physical system. One of suitable systems is 3-levels quantum state
implemented on some atom. We model the noise in such system by local spontaneous
emission channel parametrised by single real parameter
$t\in\left[0,\infty\right)$. This parameter can be understood as time. 

\subsection{Spontaneous emission channel}
Following \cite{CW2006} we chose an atom with so called V-configuration in which
the allowed spontaneous transitions are: $\ket{2} \rightarrow \ket{0}$ and
$\ket{1} \rightarrow \ket{0}$. In this analysis we assume that each atom
(qutrit) decoheres independently by the spontaneous emission. This dissipative
process is characterised by two Einstein coefficients $A_1$, $A_2$, describing
the irreversible depopulation from exited states $\ket{2}$ and $\ket{1}$. For
simplicity the following calculations are conducted with $A_1=A_2=1$.

The following set of Kraus operators represents this channel:
\begin{eqnarray*}
K_0=
\left[
\begin{smallmatrix}
 1 & 0 & 0 \\
 0 & e^{-\frac{t A_1}{2}} & 0 \\
 0 & 0 & e^{-\frac{t A_2}{2}}
\end{smallmatrix}
\right],
K_1=
\sqrt{1-e^{-t A_1}}
\left[
\begin{smallmatrix}
 0 & 1 & 0 \\
 0 & 0 & 0 \\
 0 & 0 & 0
\end{smallmatrix}
\right],
K_2=
\sqrt{1-e^{-t A_2}}
\left[
\begin{smallmatrix}
 0 & 0 & 1 \\
 0 & 0 & 0 \\
 0 & 0 & 0
\end{smallmatrix}
\right].
\end{eqnarray*}
Hence action of the channel on one qutrit is
\begin{equation}
\label{equ:se:channel}
\Phi_{SE}(\rho)=\sum_{i=0}^2 K_i\rho K_i^\dagger.
\end{equation}

The extended channel acting on all three qutrits is obtained by applying the
following formula
\begin{equation}\label{equ:channelextention}
\Phi(\rho)=\sum_{i_1,i_2, i_3=1}^3 K_{i_1}\otimes K_{i_2}\otimes K_{i_3} \rho K_{i_1}^\dagger \otimes K_{i_2}^\dagger \otimes K_{i_3}^\dagger.
\end{equation}

\subsection{Generalized Pauli channel}
Generalized Pauli channel is an extension of the Pauli channel to any dimension
\cite{Hayashi2006}. In order to apply the generalized Pauli channel on
qutrits, one defines two unitary operators:
\begin{equation}
X=
\left[
\begin{smallmatrix}
 0 & 1 & 0 \\
 0 & 0 & 1 \\
 1 & 0 & 0
\end{smallmatrix}
\right],\quad
Z=
\left[
\begin{smallmatrix}
 1 & 0 & 0 \\
 0 & e^{2/3 i \pi} & 0 \\
 0 & 0 & e^{4/3 i \pi}
\end{smallmatrix}
\right].
\end{equation} 

One can create a~family of generalized Pauli channels with the help of Kraus
operators, in the following way:
\begin{equation}
K_{i,j}=\bigcup_{i,j=0}^{2} \left\{\sqrt{P_{i,j}} X^i Y^j\right\}.
\end{equation}
By putting $P_{0,0}=1-\frac{8}{9}p$ and $P_{i,j}=\frac{1}{9}p$ for $(i,j)\neq(0,0)$ we
obtain one parameter family of noisy channels. The parameter
$p\in\left[0,1\right]$ can be understood as probability of error occurrence.
Note that for $p=1$ action of the channel transforms any state to maximally mixed state
$\1/3$.

The extension of this channel acting on three qutrits is defined in analogy to
the formula (\ref{equ:channelextention}). 

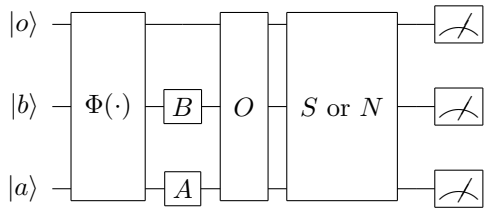
\begin{figure}
\begin{center}
\def\noise{{\Phi(\cdot)}}
\def\gateSN{{{S}{\text{ or }}{N}}}
\def\gateO{{O}}
\[ \Qcircuit @C=0.7em @R=1.7em {
    \lstick{\ket{o}} & \multigate{2}{\noise} 	& \qw & \multigate{2}{\gateO} & \multigate{2}{\gateSN} & \qw  & \meter\\
    \lstick{\ket{b}} & \ghost{\noise} 				& \gate{B} & \ghost{\gateO} 			 & \ghost{\gateSN} 			  & \qw  & \meter\\
    \lstick{\ket{a}} & \ghost{\noise}					& \gate{A} & \ghost{\gateO}				 & \ghost{\gateSN} 			  & \qw  & \meter
}\]
\caption{Graphical representation of the course of the game.}
\end{center}
\end{figure}

\section{Results}
\label{sec:results}
Flitney and Abbott considered two combinations of Bob's quantum strategies. 
One given by matrices:
\begin{equation*}
{M_1}=
\left[
\begin{matrix}
 0 & 1 & 0 \\
 0 & 0 & 1 \\
 1 & 0 & 0
\end{matrix}
\right],
\quad
{M_2}=
\left[
\begin{matrix}
 0 & 0 & 1 \\
 1 & 0 & 0 \\
 0 & 1 & 0
\end{matrix}
\right],
\end{equation*}
together with Alice's strategy $\1$ and the initial state $\ket{\Psi_1}$ (eq. \ref{equ:psi1}). 
The second with Bob's strategy $\1$ and the initial state $\ket{\Psi_2}$ (eq. \ref{equ:psi2}) and Alice's strategy
\begin{equation*}
{H}=
\left[
\begin{matrix}
 \frac{1}{\sqrt{2}} & \frac{1}{2} & \frac{1}{2} \\
 -\frac{1}{2} & \frac{3-i \sqrt{7}}{4 \sqrt{2}} & \frac{1+i \sqrt{7}}{4 \sqrt{2}} \\
 \frac{-1-i \sqrt{7}}{4 \sqrt{2}} & \frac{1}{8} \left(-3+i \sqrt{7}\right) & \frac{1}{8} \left(5+i \sqrt{7}\right)
\end{matrix}
\right].
\end{equation*} 

It should be noted that strategies $M_1$ and $M_2$ correspond to a shuffling of
Bob's choices amongst the three boxes.

We extend this analysis by sending initial pure states through quantum noisy
channel. Bob's expected pay-off $\langle\$_B\rangle$ as the function of noise
parameters $t$ or $p$ and switching parameter $\gamma$ are shown in
Fig.~\ref{fig:se} and Fig.~\ref{fig:gp}.
We have investigated the above combinations of strategies in presence of quantum
noise modelled by spontaneous emission channel and generalized Pauli channel.

\paragraph{Spontaneous emission channel}
Spontaneous emission channel defined by equations \ref{equ:se:channel} and
\ref{equ:channelextention} models the behaviour of three-level decaying atom.
Hence this model represents a~very likely physical scenario. Note that as
$t\rightarrow\infty$ this channel transforms any state into ground state
$\ket{000}$. Below we analyse in details four cases.

\subparagraph{Case 1} 
In the case of the separable state $\ket{\Psi_1}$ and trivial quantum
strategies: $A=B=\1$ Bob's expected pay-off is given by the formula
\begin{equation}
\left<\$_B \right>=\frac{1}{6} e^{-2 t} \left(3 e^{2 t}+\left(-4+8 e^t-3 e^{2 t}\right) \cos(2\gamma)\right).
\end{equation}
This case is presented in Fig.~\ref{fig:se-p1-id-id}. One can see that in this
case Bob should switch the box if the noise parameter $t<\ln(2)$. In the limit
$t\rightarrow \infty$, $\left<\$_B \right>\rightarrow \sin^2(\gamma)$ so Bob's
maximal pay-off is $1$ for $t=\infty$ and $\gamma=\pi/2$. His minimal pay-off is
$0$ for $t=\infty$ and $\gamma=0$.

\subparagraph{Case 2}
In the case of the separable state $\ket{\Psi_1}$ and following quantum
strategies: $A=\1$, $B=M_1$ or $M_2$, Bob's expected pay-off is given by the
formula
\begin{equation}
\left<\$_B \right>=\frac{1}{6} e^{-2 t} \left(3 e^{2 t}+\left(2-4 e^t+3 e^{2 t}\right) \cos(2\gamma)\right).
\end{equation}
This case is presented in Fig.~\ref{fig:se-p1-id-m1}. One can see that in this
case Bob should not switch the box. In the limit $t\rightarrow \infty$,
$\left<\$_B \right>\rightarrow \cos^2(\gamma)$ so Bob's maximal pay-off is $1$
for $t=\infty$ and $\gamma=0$. His minimal pay-off is $0$ for $t=\infty$ and
$\gamma=\pi/2$.

\subparagraph{Case 3}
In the case of entangled state $\ket{\Psi_2}$, and trivial quantum strategies
$A=B=\1$, Bob's pay-off is given by the formula
\begin{equation}
\left<\$_B \right>=\frac{1}{6} e^{-2 t} \left(3 e^{2 t}+\left(-8+8 e^t-3 e^{2 t}\right) \cos(2\gamma)\right).
\end{equation}
This case is presented in Fig.~\ref{fig:se-p2-id-id}. One can see that in this
case Bob should always switch the box. In the limit $t\rightarrow \infty$,
$\left<\$_B \right>\rightarrow \sin^2(\gamma)$ so Bob's maximal pay-off is $1$
for $t=\infty$ and $\gamma=\pi/2$. His minimal pay-off is $0$ for $t=\infty$ and
$\gamma=0$.

\subparagraph{Case 4}
In the case of entangled state $\ket{\Psi_2}$, and following quantum strategies
$A=H$, $B=\1$, Bob's expected pay-off is given by the formula
\begin{equation}
\left<\$_B \right>=\frac{1}{6}e^{-2 t} \left(3 e^{2 t}+2 \left(-1+e^t\right) \cos(2\gamma)\right).
\end{equation}
This case is presented in Fig.~\ref{fig:se-p2-gh-id}. One can see that in this
case Bob should not switch the box. In the limit $t\rightarrow \infty$,
$\left<\$_B \right>\rightarrow 1/2$. Bob's maximal pay-off is $7/12$ for
$t=\ln(2)$ and $\gamma=0$. His minimal pay-off is $5/12$ for $t=\ln(2)$ and
$\gamma=\pi/2$.

\paragraph{Generalized Pauli channel} This family of bi-stochastic channels is
interesting from quantum information point of view. It applies random unitary
rotations on the quantum state. 

\subparagraph{Case 5}
In the case of the separable state $\ket{\Psi_1}$ and quantum strategies: $A=\1=\1$ or $M_1$ or $M_2$,
Bob's expected pay-off is given by the formula
\begin{equation}
\left<\$_B \right>=\frac{1}{6} ((1-p) \cos (2 \gamma )+3-p)
\end{equation}
This case is presented in Fig.~\ref{fig:gp-p1-id-id}. One can see that in this
case Bob should switch the box if $p<1$. Bob's maximal pay-off is $2/3$
for $p=0$ and $\gamma=0$. His minimal pay-off is $1/3$ for $p=1$ and any
$\gamma$.

\subparagraph{Case 6}
In the case of the entangled state $\ket{\Psi_2}$ and quantum strategies: $A=B=\1$,
Bob's expected pay-off is given by the formula
\begin{equation}
\left<\$_B \right>=
\frac{1}{6} \left(2 p^3-4 p^2+\left(2 p^3-8 p^2+9 p-3\right) \cos (2 \gamma )+p+3\right)
\end{equation}
This case is presented in Fig.~\ref{fig:gp-p2-id-id}. One can see that Bob
should switch the box if $p>3/2-\sqrt{3}/2.$ Bob's maximal pay-off is $1$ for $p=0$ and
$\gamma=\pi/2$. His minimal pay-off is $1/3$ for $p=1$ and any $\gamma$.

\subparagraph{Case 7}
In the case of the entangled state $\ket{\Psi_2}$ and quantum strategies: $A=H$, $B=\1$,
Bob's expected pay-off is given by the formula
\begin{equation}
\left<\$_B \right>=
\frac{1}{12} \left(p^3+\left(p^2-4 p+3\right) p \cos (2 \gamma )-2 p^2-p+6\right)
\end{equation}
This case is presented in Fig.~\ref{fig:gp-p2-gh-id}. 
One can see that Bob should switch the box if $p< 1.$
Bob's maximal pay-off is ${1}/{27} \left(9+2 \sqrt{6}\right),$ 
for $p=1-\sqrt{\frac{2}{3}}$ and $\gamma=0$. 
His minimal pay-off is $1/3$ for $p=1$ and any $\gamma$.
\def\mywidth{0.47\textwidth}
\begin{figure}
\begin{center}
\subfigure[
Initial state: $\ket{\Psi_1}$, strategies: $A=\1$, $B=\1$. $\min
\langle\$_B\rangle=0$ (dark). We have $\max \langle\$_B\rangle=1$ (light). Mesh
density is 0.05.
]
{\label{fig:se-p1-id-id}\includegraphics[width=\mywidth]{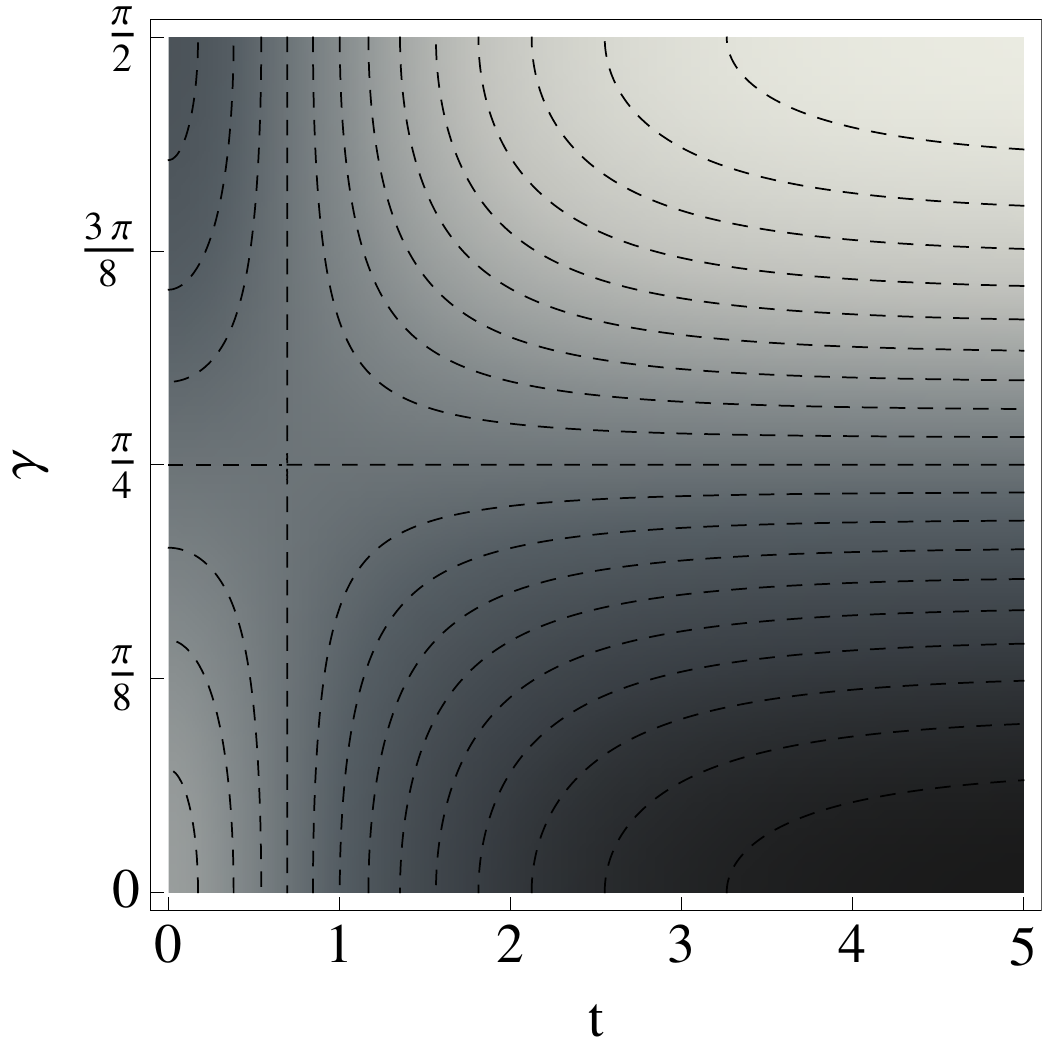}}\qquad
\subfigure[
Initial state: $\ket{\Psi_1}$, strategies $A=\1$, $B=M_1$ or $M_2$. We have
$\min \langle\$_B\rangle=0$ (dark), $\max \langle\$_B\rangle=1$ (light). Mesh
density is 0.05.
]
{\label{fig:se-p1-id-m1}\includegraphics[width=\mywidth]{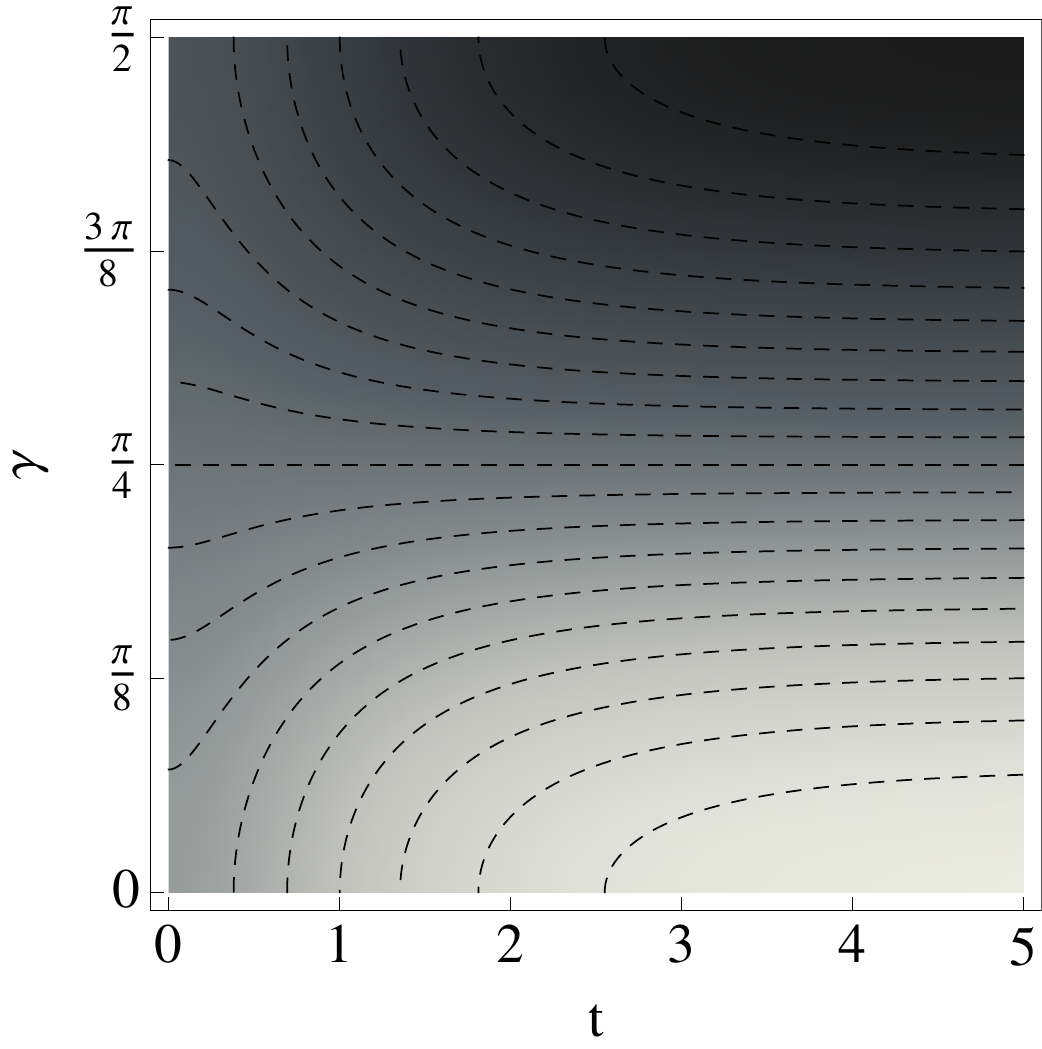}}\\
\subfigure[
Initial state: $\ket{\Psi_2}$, strategies $A=\1$, $B=\1$. We have $\min
\langle\$_B\rangle=0$ (dark), $\max \langle\$_B\rangle=1$ (light). Mesh density
is 0.05.
]
{\label{fig:se-p2-id-id}\includegraphics[width=\mywidth]{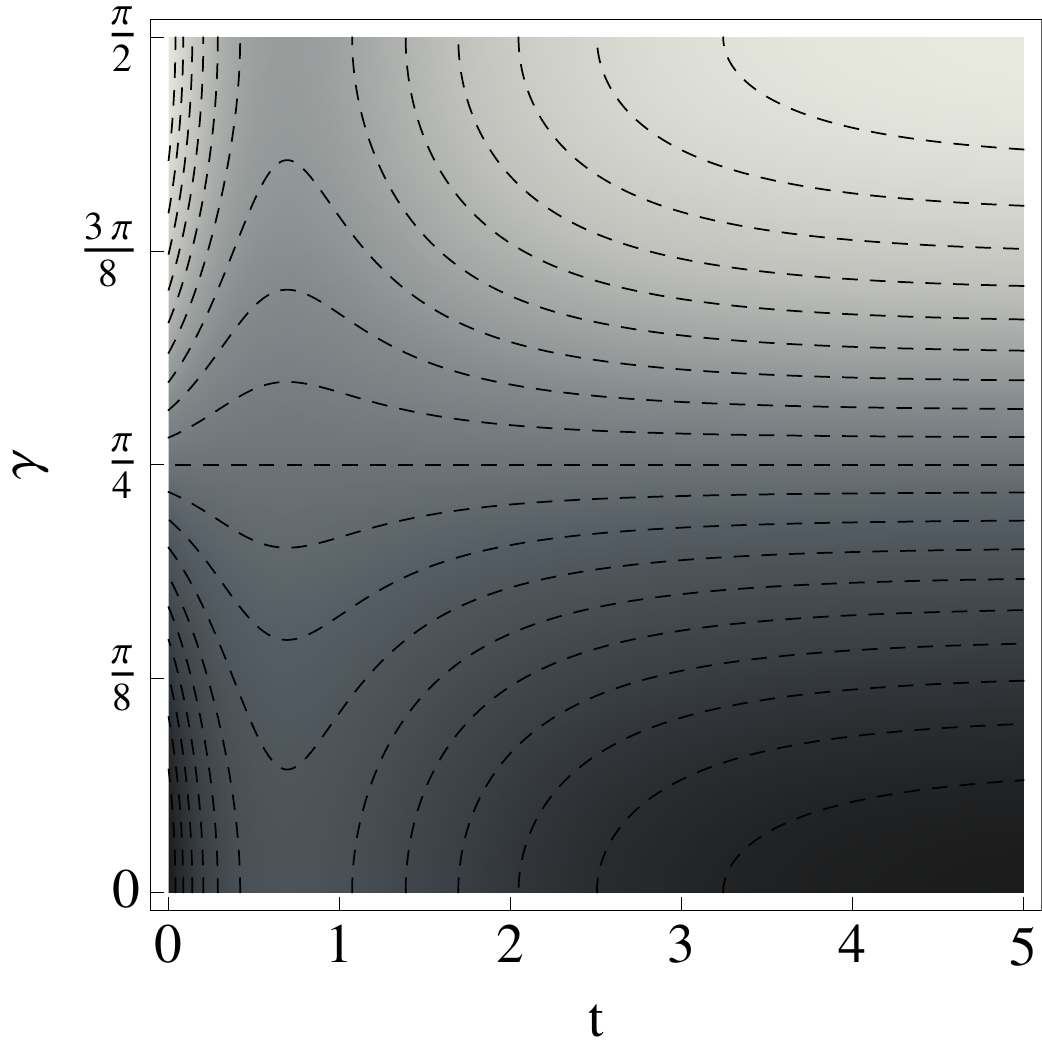}}\qquad
\subfigure[
Initial state: $\ket{\Psi_2}$, strategies $A=H$, $B=\1$. We have  $\min \langle\$_B\rangle=5/12$ (dark), $\max \langle\$_B\rangle=7/12$ (light). Mesh density is 0.01.
]
{\label{fig:se-p2-gh-id}\includegraphics[width=\mywidth]{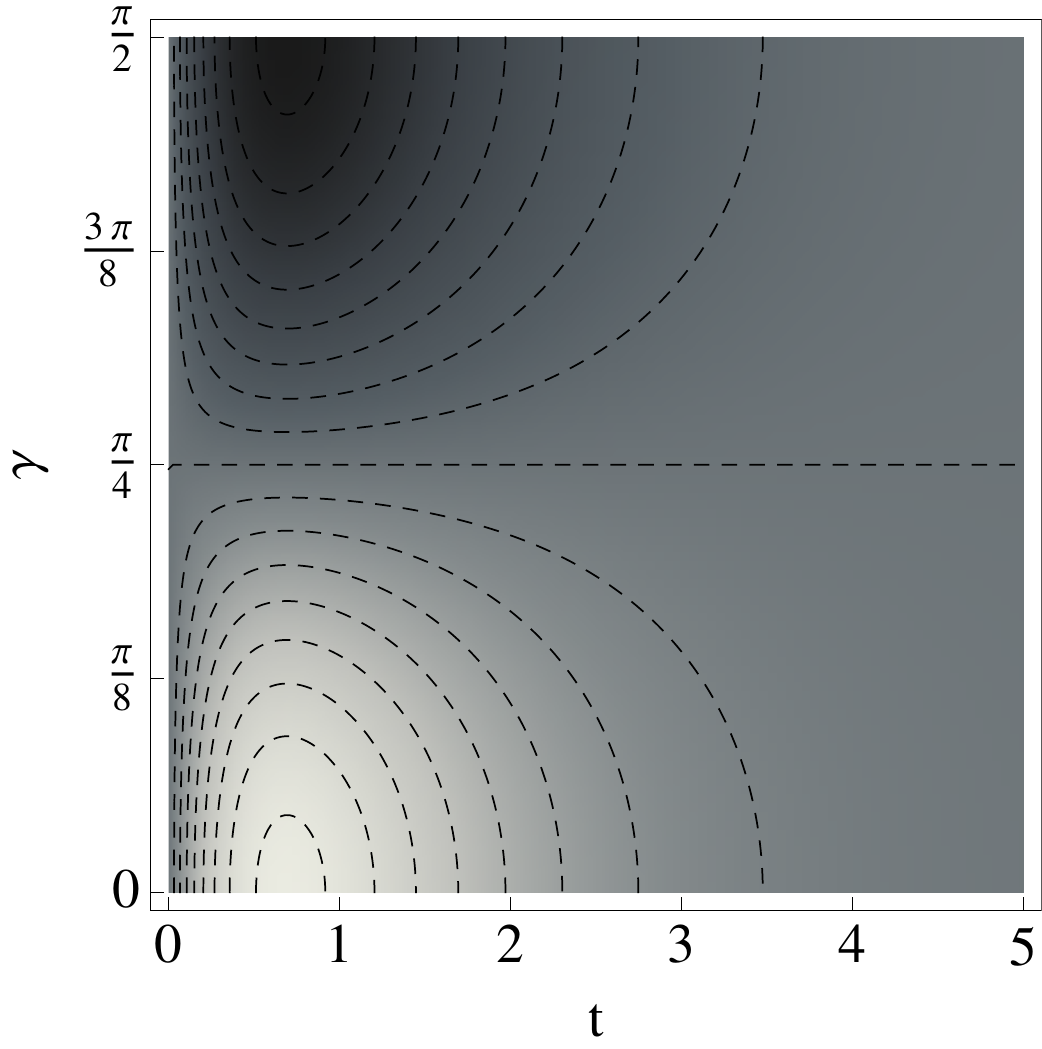}}
\caption{The behaviour on quantum Monty Hall game under the influence of
spontaneous emission channel. Bob's expected pay-off $\langle\$_B\rangle$ is
plotted as the function of noise parameter $t$ and switching parameter $\gamma$.
The colours vary from light (maximal possible pay-off) to dark (minimal
pay-off).}
\label{fig:se}
\end{center}
\end{figure}
\begin{figure}
\begin{center}
\subfigure[
Initial state: $\ket{\Psi_1}$, strategies: $A=\1$, $B=\1$ or $M_1$ or $M_2$. We
have $\min \langle\$_B\rangle=1/3$ (dark), $\max \langle\$_B\rangle=2/3$
(light). Mesh density is 0.02.
]
{\label{fig:gp-p1-id-id}\includegraphics[width=\mywidth]{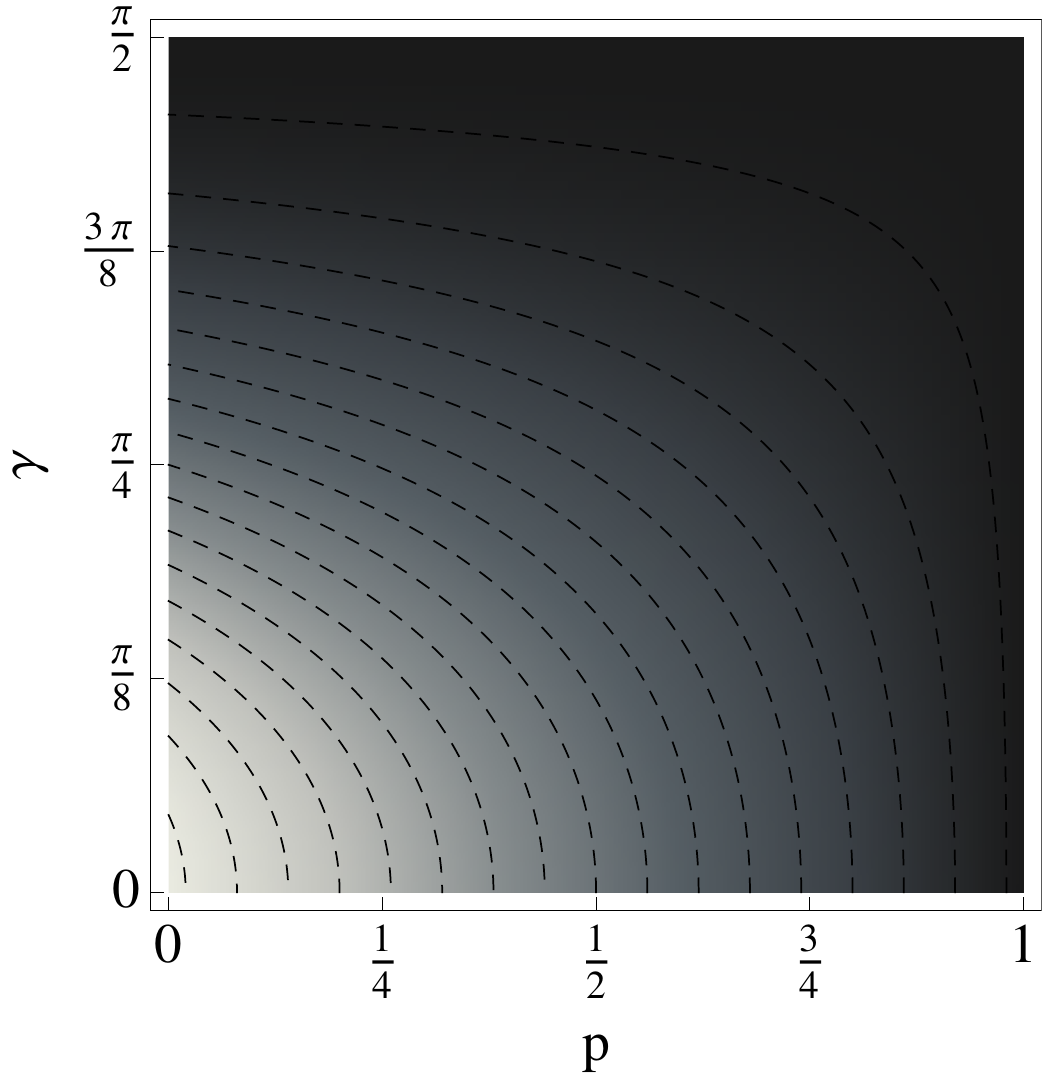}}\\
\subfigure[
Initial state: $\ket{\Psi_2}$, strategies: $A=\1$, $B=\1$. We have $\min
\langle\$_B\rangle=1/3$ (dark), $\max \langle\$_B\rangle=1$ (light). Mesh
density is 0.05. ]
{\label{fig:gp-p2-id-id}\includegraphics[width=\mywidth]{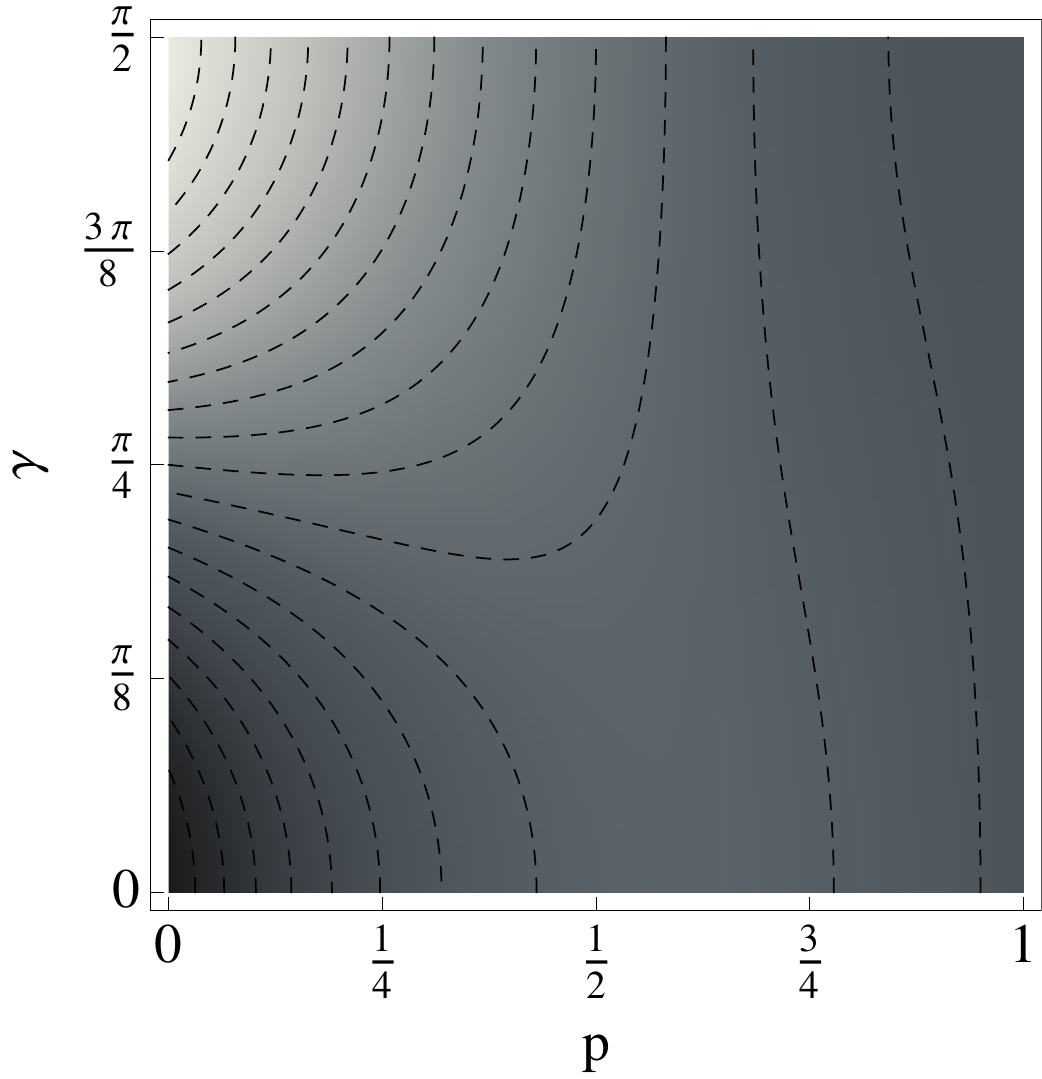}}\qquad
\subfigure[
Initial state: $\ket{\Psi_2}$, strategies: $A=H$, $B=\1$. We have $\min
\langle\$_B\rangle=1/3$ (dark), $\max \langle\$_B\rangle\approx 0.515$
(light). Mesh density is 0.01.
]
{\label{fig:gp-p2-gh-id}\includegraphics[width=\mywidth]{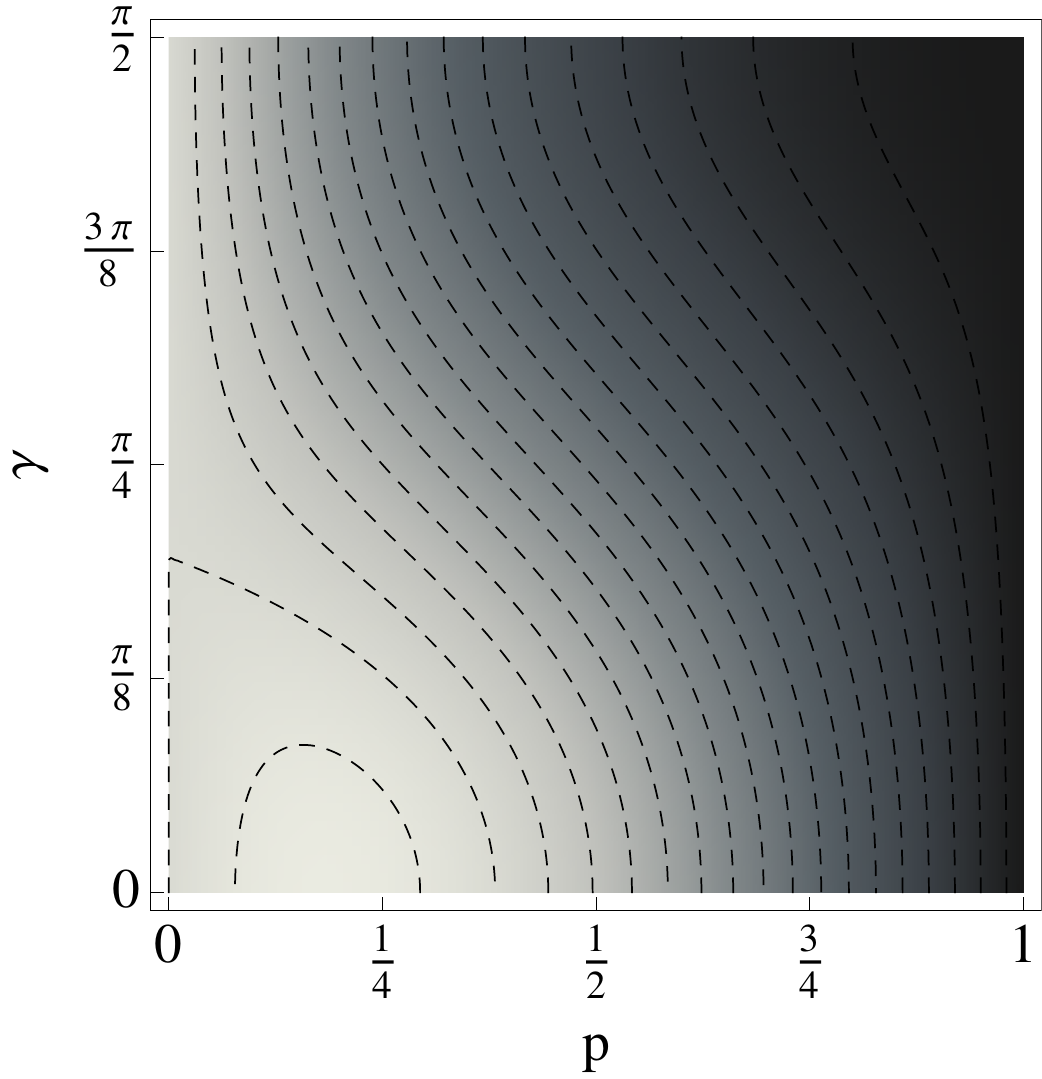}}
\caption{The behaviour on quantum Monty Hall game under the influence of
generalized Pauli channel. Bob's expected pay-off $\langle\$_B\rangle$ is
plotted as the function of noise parameter $p$ and switching parameter $\gamma$.
The colours vary from light (maximal possible pay-off) to dark (minimal
pay-off).}
\label{fig:gp}
\end{center}
\end{figure}
\paragraph{Conclusions}
We have studied two noise models, that are likely to occur in the physical
implementation of the quantum version of the Monty Hall game.

The calculation have shown that existence of noise heavily influences the
expectation value of the quantum Monty Hall game. We can observe different
behaviour of the game for different strategies, initial states and quantum
channels.

In the case of spontaneous emission channel the initial state of the game
approaches the ground state $\ket{000}$ as noise parameter goes to infinity.
Therefore for large amounts of noise the game is played (quantum strategies are
applied) on the state $\ket{000}$ rather than on states $\ket{\Psi_1}$ or
$\ket{\Psi_2}$. When we compare the asymptotic behaviour of the pay-off
functions in the cases 1 and 3 we see that they converge to the same limit
$\sin^2\gamma$. The difference between those cases is that in case 1 Bob should
switch the gate only if noise parameter is larger than $\ln(2)$.

It should be noted that for $t=\infty$ Bob knows exactly what is the initial
state of the game --- where the prize is hidden. Therefore given Alice will not
act (her strategy is to apply an identity), Bob can allways win and the game
becomes unfair. This can be observed in cases 1, 2 and 3.

Generalized Pauli channel transforms any input state towards the maximally mixed
state. Therefore for maximal value of noise parameter any correlations are lost.
It can be easily seen that if initial state of this game is maximally mixed
then, Bob's pay-off is always equal to $1/3$ and is independent of Alice's and
Bob's strategies. The noise can influence the outcomes of the game in
a~non-trivial way. In example in case 6 Bob should change his strategy, to
switching the box, when noise parameter is larger than $3/2-\sqrt{3}/2.$

Obtained results show, that if Bob knows the properties of the noise in the
system implementing quantum Monty Hall game he can use this knowledge to change
his strategy in order to maximize his pay-off. In some cases the noise can be
more influential than strategies and therefore can impede successful
implementation of quantum games. More careful studies, that take into account
imperfections of the physical device realizing this game, would be needed if
a~proposition of concrete physical implementation of this game will appear. 

\section*{Acknowledgements}
This research was partially supported by the Polish Ministry of Science and
Higher Education project N519 012 31/1957. I~would like to thank Jaros\l{}aw
Miszczak, Jan S\l{}adkowski and an anonymous referee for careful reading of the
manuscript and helpful suggestions. 

\end{document}